\documentclass[pdflatex,sn-mathphys-num]{sn-jnl}

\usepackage{graphicx}%
\usepackage{multirow}%
\usepackage{amsmath,amssymb,amsfonts}%
\usepackage{amsthm}%
\usepackage{mathrsfs}%
\usepackage[title]{appendix}%
\usepackage{xcolor}%
\usepackage{textcomp}%
\usepackage{manyfoot}%
\usepackage{booktabs}%
\usepackage{algorithm}%
\usepackage{algorithmicx}%
\usepackage{algpseudocode}%
\usepackage{listings}%
\usepackage{tabularx}

\theoremstyle{thmstyleone}%
\theoremstyle{thmstyletwo}%

\theoremstyle{thmstylethree}%

\begin{document}

\title[Article Title]{\textsc{LQCDMaster}: Agentic Scientific Computing for Lattice Quantum Chromodynamics Research}


\author[1,2,3]{\fnm{Haofei} \sur{Gao}}
\equalcont{These authors contributed equally to this work.}

\author[4,5]{\fnm{Tingjia} \sur{Miao}}
\equalcont{These authors contributed equally to this work.}

\author[4,7]{\fnm{Wenkai} \sur{Jin}}
\author[1,2,3]{\fnm{Muhua} \sur{Zhang}}
\author[1,2,3]{\fnm{Hanzhang} \sur{Wang}}
\author[1,2,3]{\fnm{Jie} \sur{Ran}}
\author[1,2,3]{\fnm{Jinxin} \sur{Tan}}
\author[1,5]{\fnm{Zhentao} \sur{Zhang}}
\author[1,5]{\fnm{Bo} \sur{Tang}}
\author[1,2,3]{\fnm{Leiyi} \sur{Li}}
\author[8]{\fnm{Jun} \sur{Hua}}
\author[9]{\fnm{Xiangyu} \sur{Jiang}}

\author*[6]{\fnm{Qi'an} \sur{Zhang}}\email{zhangqa@buaa.edu.cn}
\author*[4,5]{\fnm{Siheng} \sur{Chen}}\email{sihengc@sjtu.edu.cn}
\author*[1,2,3]{\fnm{Wei} \sur{Wang}}\email{wei.wang@sjtu.edu.cn}

\affil[1]{\orgdiv{School of Physics and Astronomy}, 
          \orgname{Shanghai Jiao Tong University}, 
          \orgaddress{\state{Shanghai}, 
                      \postcode{200240}, 
                      \country{China}}}

\affil[2]{\orgdiv{Tsung-Dao Lee Institute}, 
          \orgname{Shanghai Jiao Tong University}, 
          \orgaddress{\state{Shanghai}, 
                      \postcode{200240}, 
                      \country{China}}}

\affil[3]{\orgdiv{State Key Laboratory of Dark Matter Physics}, 
          \orgname{Shanghai Jiao Tong University}, 
          \orgaddress{\state{Shanghai}, 
                      \postcode{200240}, 
                      \country{China}}}

\affil[4]{\orgdiv{School of Artificial Intelligence}, 
          \orgname{Shanghai Jiao Tong University}, 
          \orgaddress{\state{Shanghai}, 
                      \postcode{200030}, 
                      \country{China}}}

\affil[5]{\orgdiv{Zhiyuan College}, 
          \orgname{Shanghai Jiao Tong University}, 
          \orgaddress{\state{Shanghai}, 
                      \postcode{200240}, 
                      \country{China}}}

\affil[6]{\orgdiv{School of Physics}, 
          \orgname{Beihang University}, 
          \orgaddress{\state{Beijing}, 
                      \postcode{102206}, 
                      \country{China}}}

\affil[7]{\orgname{SciLand}, 
          \orgaddress{\state{Shanghai}, 
                      \postcode{200030}, 
                      \country{China}}}

\affil[8]{\orgdiv{Institute of Quantum Matter}, 
          \orgname{South China Normal University}, 
          \orgaddress{\state{Guangzhou}, 
                      \postcode{510006}, 
                      \country{China}}}

\affil[9]{\orgdiv{Department of Physics}, 
          \orgname{Indiana University}, 
          \orgaddress{\state{Bloomington}, 
                      \postcode{Indiana 47405}, 
                      \country{USA}}}


\abstract{
Lattice quantum chromodynamics (LQCD) provides a first-principles framework for computing hadronic observables, but its practical use remains limited by the substantial expertise required to turn research motivation into reliable computing workflows. Here we present \textsc{LQCDMaster}, a tool-augmented, skill-guided and domain-specialized scientific computing agent that converts natural-language LQCD research tasks into executable PyQUDA computing workflows, including measurement scripts, job-submission artifacts, execution logs and numerical outputs. The system combines agentic planning, expert-annotated LQCD skills and a deterministic Wick-contraction tool to constrain the algebraically fragile components of code generation. We evaluate \textsc{LQCDMaster} on a benchmark at the forefront of scientific research, comprising 70 LQCD computing tasks, with observables covering local and nonlocal two-point functions, Wilson loops, meson and baryon three-point functions. The generated workflows exactly reproduce expert-written implementations in 63 of 70 tasks at machine precision, with three additional discrepancies attributable to convention mismatches. Across representative observables, the agent reduces implementation time from hours to minutes while preserving end-to-end numerical validation. Further, we present a typical case of \textsc{LQCDMaster}-driven exploration: a lattice computation of light-cone distribution amplitudes with diagonal Wilson-line, a quantity accessible with standard methods but never before computed, and computation of the spectrum of proton, deuteron, triton, hyperon ($\Lambda$), hyperdeuteron ($p\Lambda$) and  hypertriton ($pn\Lambda$). This work pioneers the paradigm of agentic scientific computing by automating the end-to-end scientific computing workflows in lattice QCD research, lowering its barrier and facilitating the exploration and verification of non-standard scientific ideas. 
}




\maketitle

\section{Introduction}

\subsection{The Barrier to Entry in Lattice QCD}

Lattice quantum chromodynamics (LQCD) provides a first-principles definition of the strong nuclear force directly from the QCD Lagrangian~\cite{Wilson:1974sk}. Over the past four decades, it has matured into a precision tool for calculating hadron masses~\cite{BMW:2008jgk,BMW:2014pzb}, studying nucleon structure~\cite{Chang:2018uxx}, and probing new physics~\cite{Borsanyi:2016ksw,Boccaletti:2024guq}. A typical LQCD workflow proceeds through several stages: defining the physics observable, designing the computational scheme, writing code in libraries such as Chroma~\cite{Edwards:2004sx} or PyQUDA~\cite{Jiang:2024lto}, debugging on test configurations, scaling to production on GPU clusters, and analyzing the resulting correlators.

 Among these stages, converting a physics idea into a runnable script is the most challenging. It requires simultaneous expertise in LQCD physics (Wick contractions, $\gamma$-matrix algebra, parity projectors), API conventions (gauge context management, lattice field layouts, MPI reduction), and numerical methods (multi-grid solvers, gauge smearing). Human experts typically require from hours to days per observable, and sometimes longer, with three to five debugging cycles before code can be trusted in production.

 A subtler but equally important barrier is cognitive: the high engineering cost of implementing a new idea discourages exploration. Many physically interesting but non-standard computations, such as those involving different $\gamma$-matrix combinations, unconventional source-sink separations, and systematic scans over operator definitions, are conceptually trivial but practically prohibitive. As a result, such computations are often not attempted, producing a form of \textbf{exploration bias} in the LQCD literature: only the most conventional observables are computed, not because other ideas lack physical merit, but because they lack engineering justification.

\subsection{Related Work}

Machine learning has already become a useful numerical ingredient in lattice QCD. Existing work primarily uses models within established computational pipelines, for example to accelerate gauge-field generation through flow-based sampling~\cite{Albergo:2019eim,Kanwar:2020xzo,Wang:2023exq,Aarts:2025gyp} or to learn effective actions. In such cases, AI serves as a numerical tool, but the surrounding workflow remains human-centered.

Beyond AI as a numerical tool, LLM-based agents have demonstrated further capabilities in automating real scientific research. Agentic systems like Robin and Kosmos integrate literature retrieval, experimental iteration, and result reporting, realizing an end-to-end research workflow ~\cite{ghareeb2025robin, mitchener2025kosmos}. Meanwhile, the diversity of scientific domains has motivated the development of powerful domain-specialized agents. For example, ML-Master demonstrates expert-level performance on machine-learning research tasks, highlighting the potential of AI for AI~\cite{liu2025ml}. \textsc{LQCDMaster} further extends the paradigm of agentic scientific research to Lattice QCD, a frontier domain in high-energy physics with a high knowledge barrier.

General-purpose code-generation systems such as Codex~\cite{codex} and Claude Code~\cite{claude} can accelerate software drafting, but are not designed to guarantee the algebraic and field-theoretic correctness required by lattice computations. In LQCD, a program may be syntactically valid, execute without runtime errors, and nevertheless produce an invalid observable because of an incorrect spin--color contraction, flavor assignment, or fermionic sign. The central challenge for scientific program generation is therefore generating executable workflows that are grounded in domain theory, constrained by expert-defined skills, and coupled to deterministic tools for symbolic manipulation.

\subsection{Contributions of This Work}

We present \textsc{LQCDMaster}, \textbf{a tool-augmented, skill-guided, domain-specialized} scientific computing agent for LQCD code generation. Its core contributions are as follows:

\begin{enumerate}
\item \textbf{Lowering the barrier to LQCD research}: \textsc{LQCDMaster} converts natural-language research tasks (e.g., {\it Calculate the two-point function of a pion with the operator $\bar{u}\gamma_5 d$. Use the point source at position [0,0,0,0].}) into executable LQCD computing workflows, allowing physicists to start LQCD research without manually writing implementation code. This reduces the required implementation workload, accelerates research, and broadens access to this technically demanding frontier of high-energy and nuclear physics.

\item \textbf{Providing a comprehensive LQCD benchmark and rigorous evaluation}. We construct a comprehensive LQCD coding benchmark comprising 70 representative tasks spanning five observable classes, all drawn from the forefront of scientific research. We conduct a rigorous quantitative evaluation. Using GPT-5.4 as the backbone model, our system successfully reproduces the human-expert implementation results for 63 of the 70 tasks at machine precision. The remaining cases exhibit either convention mismatches, such as phase mismatches, or algorithmic and numerical errors.

\item \textbf{Pioneering the paradigm of agentic scientific computing}: \textsc{LQCDMaster} automates end-to-end scientific computing workflows in lattice QCD research with expert-level accuracy, while substantially improving efficiency and reducing the per-observable implementation cost from hours to minutes. As concrete case studies, we present the first lattice computations of LCDAs using diagonal Wilson lines, which are accessible with standard methods but have not previously been computed, and multi-hadron spectroscopy, for which Wick-contraction complexity grows rapidly with hadron number. \textsc{LQCDMaster} automates these studies, paving a viable path to agentic scientific research through human--AI collaboration.

\end{enumerate}


\section{Results}
\label{sec:results}

\subsection{Benchmarking End-to-end Lattice QCD Code Generation}
\label{subsec:end-to-end-benchmark}

\textsc{LQCDMaster} is an autonomous scientific computing agent for lattice QCD. Given a natural-language research request specifying the target observable, ensemble and kinematic setup, the system produces a reviewable PyQUDA workflow comprising the measurement script, job-submission artifacts, execution logs and numerical outputs. We evaluated this workflow-generation capability on a benchmark suite spanning representative lattice QCD production tasks, with the aim of testing whether the generated workflows reproduce expert-written reference implementations at numerical precision while retaining comparable computational performance.

The benchmark suite consisted of 70 independent tasks on the C24P29 ensemble generated by the CLQCD collaboration~\cite{CLQCD:2023sdb,Zhang:2021oja}, with lattice spacing $a=0.105~\mathrm{fm}$ and lattice volume $24^3\times72$. The pion mass on this ensemble is about $290~\mathrm{MeV}$. The tasks were selected to cover qualitatively distinct structures in lattice QCD computing: they included 20 local two-point functions (2PTs) for mesons and baryons, 10 nonlocal Wilson-line 2PTs, 13 meson three-point functions (3PTs), 15 baryon 3PTs using sequential sources, and 12 Wilson-loop measurements. Together, these workflows probe the main implementation challenges encountered in production LQCD code, including spin--color contractions, flavor assignments, gauge-link path ordering, source--sink kinematics, sequential-propagator conventions, MPI-parallel data layout, and output normalization.

\begin{table}[t]
\centering
\caption{
Representative Lattice QCD workflow categories in the benchmark.
The table summarizes the computational structure that \textsc{LQCDMaster} must recover
from natural-language task specifications. The listed challenges are not exhaustive;
they identify the dominant implementation difficulty for each observable class.
}
\label{tab:lqcd-task-overview}
\small
\renewcommand{\arraystretch}{1.18}
\setlength{\tabcolsep}{4.5pt}
\begin{tabular}{lp{4.2cm}p{6cm}}
\toprule
\textbf{Observable} & \textbf{Representative workflow} & \textbf{Challenges} \\
\midrule
Local 2pt
& Meson and baryon two-point correlators with local source and sink operators
& Correct construction of spin--color contractions, including meson traces and baryon $\epsilon$-tensor structures. \\

Nonlocal 2pt
& Gauge-invariant two-point correlators of spatially separated quark fields
& Gauge-covariant transport between displaced fields, together with a displacement-dependent phase. \\

Wilson loop
& Pure-gauge closed-loop observables on specified contours in space and time
& Path ordering, loop geometry, orientation conventions and gauge-link multiplication along closed contours. \\

Meson 3pt
& Three-point correlators of mesons with a local current insertion
& Sequential-source construction, sink projection, current insertion. \\

Baryon 3pt
& Three-point correlators of baryons with three-quark interpolating operators and a current insertion
& Spin--color routing through baryon contractions, transition-specific index structure and consistency of source, sink and insertion conventions. \\
\bottomrule
\end{tabular}
\end{table}

All generated scripts in the main benchmark were syntactically valid and completed test execution. Because successful execution does not by itself establish scientific correctness, each generated workflow was validated by direct numerical comparison with a hand-written reference implementation for the same observable, ensemble and kinematic setup. This comparison provides an end-to-end test of whether the generated program encodes the intended Wick contractions, $\gamma$-matrix structures, flavor channels and normalization conventions, rather than merely producing executable Python code.

\subsection{Expert-level Accuracy with Improved Implementation Efficiency}
\label{subsec:numerical-agreement}

\begin{figure}[htbp]
\centering
\includegraphics[width=0.8\linewidth]{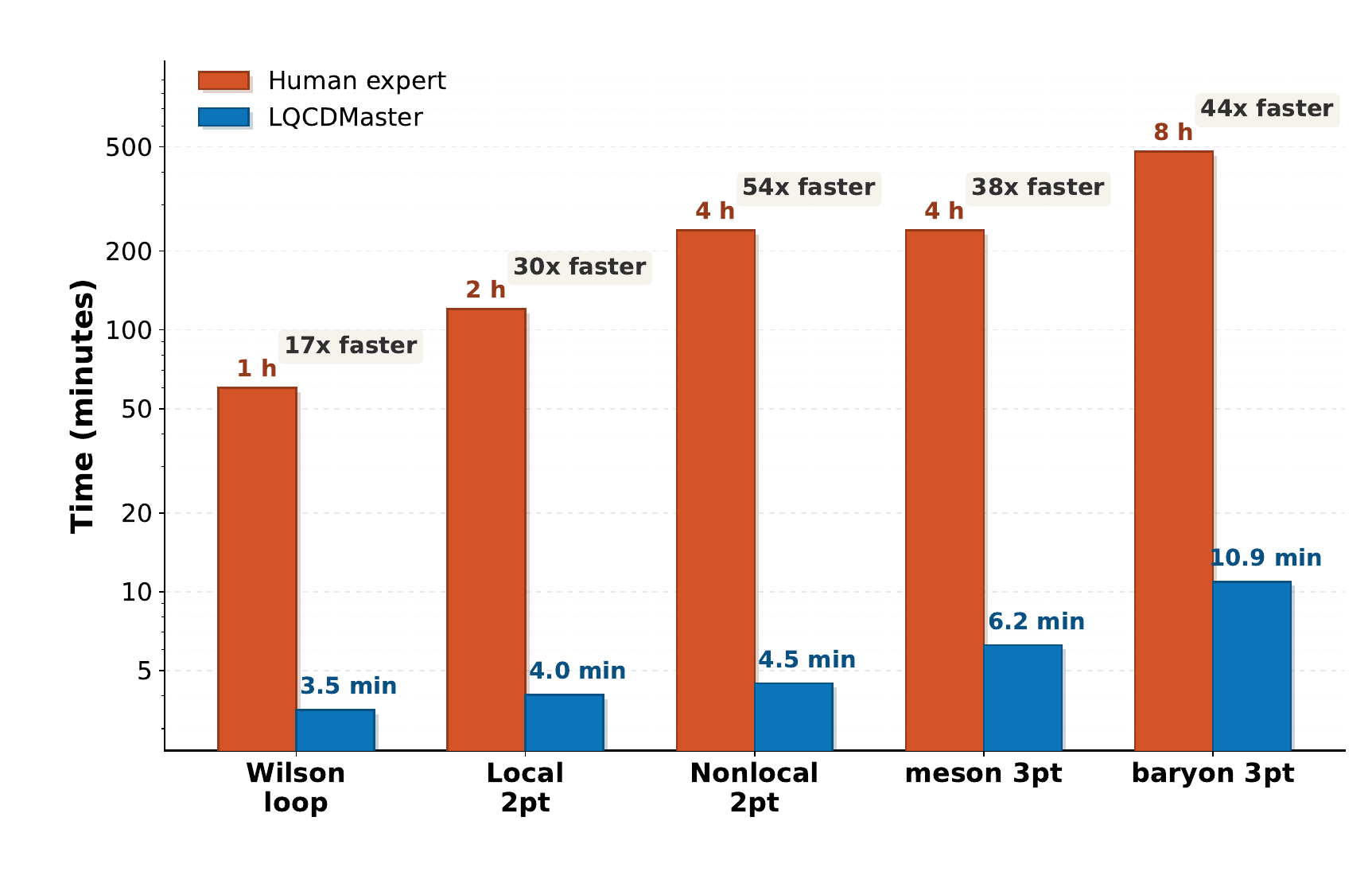}
\caption{Implementation-efficiency improvements across representative lattice QCD computing tasks, comparing the time required for human experts and \textsc{LQCDMaster} to produce reliable computing workflows.}
\label{fig:development-time}
\end{figure}

\begin{table}[htbp]
\centering
\caption{
Expert-level accuracy of \textsc{LQCDMaster}-generated lattice QCD computing workflows and alignment with expert-written implementations, with GPT-5.4 as the backbone model.
}
\label{tab:cross-validation}
\small
\renewcommand{\arraystretch}{1.15}
\setlength{\tabcolsep}{5.0pt}
\begin{tabular}{lccccc}
\toprule
\textbf{Observables} & \textbf{Tasks} & \textbf{Matched} & \textbf{Convention Mismatch} & \textbf{Failure} & \textbf{Accuracy} \\
\midrule
Local 2pt              & 20 & 20 & 0 & 0 & 100.0\% \\
Nonlocal 2pt           & 10 & 8  & 2 & 0 & 80.0\%  \\
Wilson loop            & 12 & 12 & 0 & 0 & 100.0\% \\
Meson 3pt              & 13 & 13 & 0 & 0 & 100.0\% \\
Baryon 3pt             & 15 & 10 & 1 & 4 & 66.7\% \\
\midrule
\textbf{Total}         & \textbf{70} & \textbf{63} & \textbf{3} & \textbf{4} & \textbf{90.0\%} \\
\bottomrule
\end{tabular}

\vspace{0.5em}
\begin{minipage}{0.95\linewidth}
\footnotesize
\begin{itemize}
    \setlength{\itemsep}{0.2em}
    \setlength{\leftmargin}{1.2em}
    \item \textbf{Matched}: numerical agreement with the expert-written reference implementation at machine precision, $|\Delta|\lesssim10^{-12}$, under identical input data, normalization choices and output conventions.
    \item \textbf{Convention mismatch}: minor discrepancy like single global sign or phase transformation, with the full correlator structure preserved after fixed convention alignment.
    \item \textbf{Failure}: any remaining algebraic or numerical discrepancy after convention alignment.
\end{itemize}
\end{minipage}
\end{table}

\textsc{LQCDMaster} achieved high implementation-level accuracy on the 70-task benchmark (Table~\ref{tab:cross-validation}). Across local and nonlocal two-point functions, Wilson loops, meson three-point functions and baryon three-point functions, 63 generated workflows reproduced expert-written reference implementations at machine precision, with $|\Delta|\lesssim10^{-12}$. The remaining seven cases comprised three convention mismatches and four unresolved failures. The three convention mismatches, two in nonlocal two-point functions and one in baryon three-point functions, were traced to global sign conventions: after applying a single overall sign transformation, the full time- and displacement-dependent correlators agreed with the reference. The four unresolved failures involved baryon three-point functions.

The benchmark probes several distinct implementation burdens in lattice QCD measurements, including meson traces, baryon $\epsilon$-tensor contractions, gauge-link transport, Wilson-loop path ordering, sequential-source construction and spin--color routing in baryon transition amplitudes. The results therefore evaluate more than executable code generation; they test whether the agent can instantiate the domain-specific computational structure required by heterogeneous observable classes. This distinction is essential in lattice QCD, where a single index, phase or contraction error can produce plausible but incorrect correlators. Here, numerical validation converted the observed discrepancies into explicit convention-level outcomes rather than leaving them as silent implementation failures.

Further, we assess the extent to which \textsc{LQCDMaster} reduces the implementation time required for lattice QCD computing. We therefore compared the per-observable time required by lattice QCD experts with the time required by \textsc{LQCDMaster} to generate and verify the corresponding workflow (Fig.~\ref{fig:development-time}). Expert implementation time ranged from 1 hour for Wilson loops to 8 hours for baryon three-point functions, with local two-point functions requiring 2 hours and nonlocal two-point and meson three-point functions requiring 4 hours each. \textsc{LQCDMaster} completed the corresponding workflows in 3.5, 4.0, 4.5, 6.2 and 10.9 minutes, respectively, corresponding to speedups of 17$\times$, 30$\times$, 54$\times$, 38$\times$, and 44$\times$. Overall, the speedup tends to increase with the complexity of the observable, suggesting that \textsc{LQCDMaster} provides more substantial time savings for computationally demanding tasks.

 The four tasks that failed numerical comparison in that run involved specific transition-current combinations, such as the $p \to p$ vector current, the $\Lambda \to \Lambda$ vector and axial currents, and the $\Lambda_b \to \Lambda_c$ vector current. They passed preliminary static checks and test execution but failed at the numerical-comparison stage. These cases represent the primary failure mode that the current critique--rewrite pipeline did not fully catch.

These gains reflect a reduction in marginal implementation effort, not a change in the underlying lattice QCD algorithms. Once a workflow is generated, production measurements remain governed by solver cost, ensemble size and gauge-configuration statistics. The largest gains arise when a task contains substantial convention, contraction and debugging overhead that can be absorbed by deterministic tools, reusable domain skills and automated numerical checks.

\subsection{New-observable Exploration Driven by \textsc{LQCDMaster}}
\label{subsec:diagonal-lcda}

\begin{figure}[htbp]
  \centering
\includegraphics[width=1\linewidth]{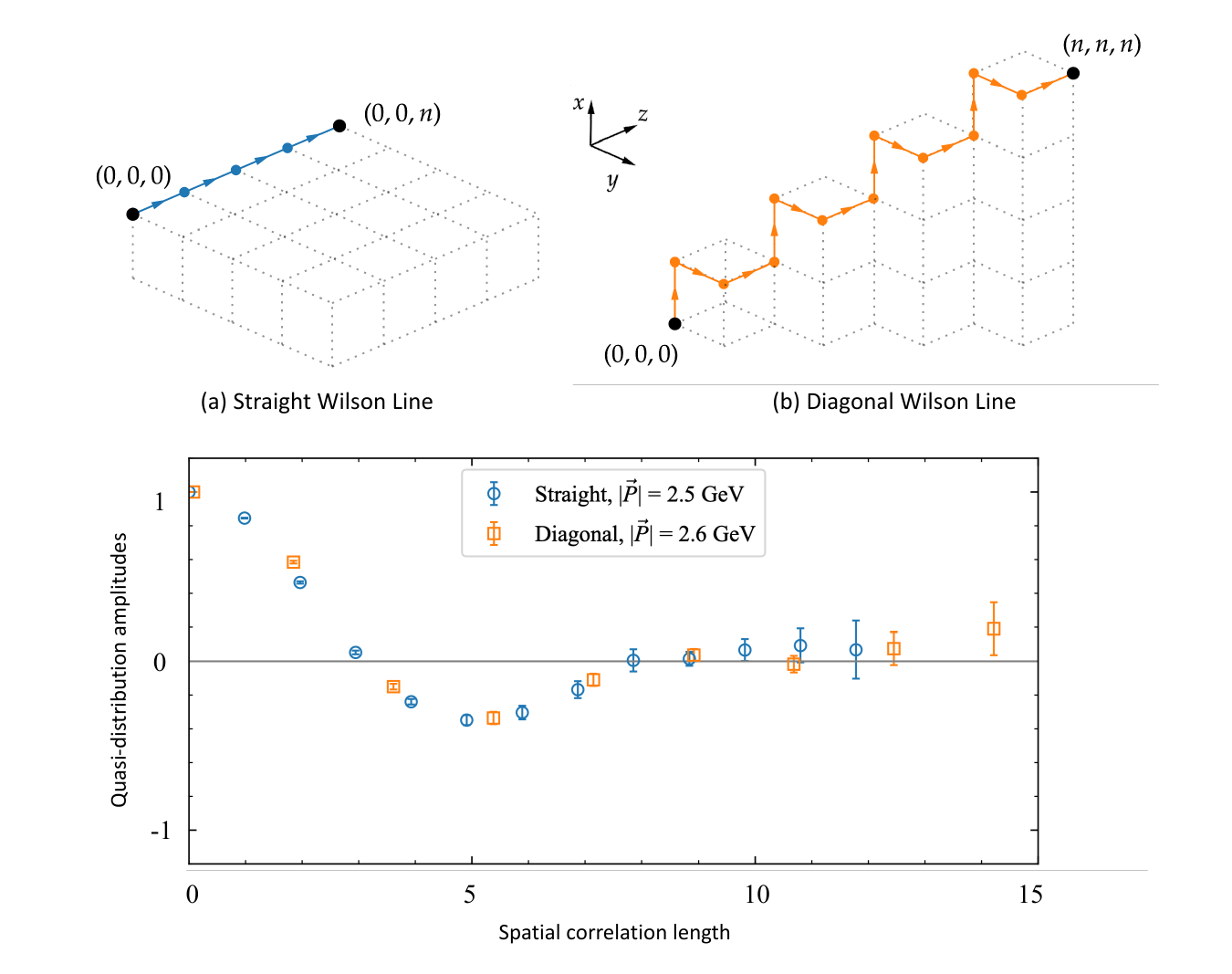}
  \caption{Physical schematic of the meson LCDA setup with a non-standard diagonal Wilson line shown in the upper panels. The real parts of the matrix elements are shown in the lower panel. }
  \label{fig:diagonal_LCDA}
\end{figure}

\begin{figure}[htbp]
  \centering
\includegraphics[width=0.85\linewidth]{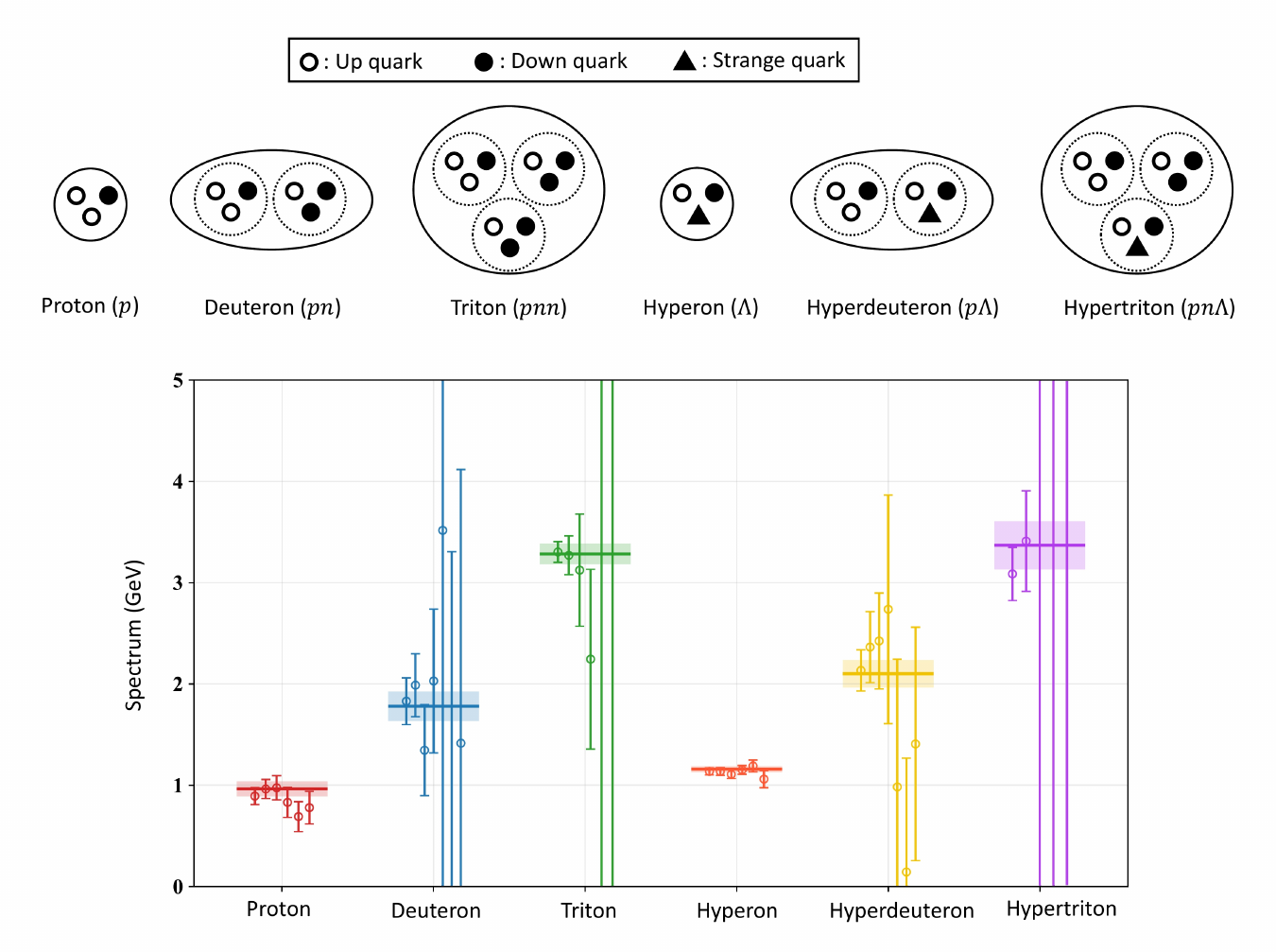}
  \caption{Multi-hadron mass results generated with \textsc{LQCDMaster}. In contrast to conventional manual workflows, which become rapidly intractable as contraction complexity grows with hadron number, the agentic workflow keeps the implementation practical.}
  \label{fig:multi_hadron_mass}
\end{figure}

The benchmark suite tests whether \textsc{LQCDMaster} can automate established workflows. A key question is whether lowering the implementation cost changes which scientific ideas are practical to explore. To test this, we applied the system to a non-standard but well-defined measurement: a meson light-cone distribution amplitude constructed with a diagonal Wilson-line geometry.

Conventional lattice computations of quasi-distribution amplitudes~\cite{Ji:2013dva} usually employ straight Wilson lines along a spatial lattice axis. Diagonal Wilson lines provide an alternative geometry that enlarges the accessible set of path directions and effective kinematics on a fixed ensemble. The idea is straightforward at the level of the continuum operator, but implementing it by hand requires modifying gauge-link transport, displacement bookkeeping, phase conventions and nonlocal contractions as shown in Fig.~\ref{fig:diagonal_LCDA}. Under a conventional workflow, this cost makes such variants easy to postpone unless the expected physics return is already clear.

With \textsc{LQCDMaster}, the diagonal Wilson-line measurement was specified as a natural-language request and converted into executable PyQUDA code within minutes. The generated workflow was then cross-checked against an independently written PyQUDA implementation on the same ensemble and agreed within statistical uncertainties. The resulting real parts of the reduced matrix element are shown in Fig.~\ref{fig:diagonal_LCDA}. To our knowledge, this is the first lattice QCD evaluation of a meson distribution-amplitude observable using a diagonal Wilson-line geometry.

A second representative case is multi-hadron spectroscopy, including proton, deuteron and  triton,   the hyperon $\Lambda$, hyperdeuteron $p\Lambda$ and hypertriton  $pn\Lambda$ systems. Spectra and interactions of the hypernuclear systems are both extremely important in the formation of neutron stars\cite{Tolos:2020aln}. 
In conventional hand-written implementations, this class of tasks is difficult mainly because the contraction complexity grows rapidly with hadron number: as the operator basis grows, the number of Wick-contraction topologies increases:  there are   $2$, $36$ and 2880, 1, 12, and 576 contraction terms in the 2PTs. Accordingly, the manual derivation, coding and debugging become increasingly error-prone. Using \textsc{LQCDMaster},  we have  generated the corresponding workflow from a natural-language specification within several minutes. The obtained results for the mass of   multi-hadron systems are shown  in Fig.~\ref{fig:multi_hadron_mass}.

The significance of these examples is not that the agent independently discovered new operators. The ideas came from human physics reasoning. The contribution of the computational agent was to reduce the engineering cost of testing such ideas to the point where they became worth trying. This illustrates a broader mechanism for AI-assisted computational science: reliable workflow generation can expand scientific exploration by lowering the marginal cost of implementing non-standard but theoretically motivated observables.

\subsection{Failure Modes Analysis}
\label{subsec:failure-analysis}

A scientific code-generation agent is useful only if its errors can be exposed before production use. We therefore analyzed the first-pass issues observed during the solve--critique--rewrite process and the subsequent numerical validation (Table~\ref{tab:failure-modes}). Most issues were conventional implementation errors, including einsum-index mismatches, gauge-context handling mistakes, output-format inconsistencies and array-shape mismatches. These errors were detected by static checks, test execution or automated critique, and were corrected before final validation.

The remaining failure mode was more domain-specific. Several baryon three-point workflows passed syntax checks and test execution but failed direct numerical comparison with expert-written references. These cases involved transition-current combinations, such as the $p\to p$ vector current, the $\Lambda\to\Lambda$ vector and axial currents and the $\Lambda_b\to\Lambda_c$ vector current. They indicate that executable code and static consistency checks are insufficient for validating baryon contraction logic; direct numerical regression against trusted references remains necessary for complex spin--color routing and transition-specific conventions.

To conclude, \textsc{LQCDMaster} reliably removes most routine implementation errors through automated critique, checks and executable tests. Meanwhile, for new observable classes and other novel tasks, reference comparisons and human review remain essential components of a robust scientific computing workflow.

\begin{table}[htbp]
\centering
\caption{
Observed failure modes during \textsc{LQCDMaster} workflow generation and validation. Most routine issues were corrected automatically, whereas the remaining failures required direct numerical comparison with expert-written references.
}
\label{tab:failure-modes}
\small
\renewcommand{\arraystretch}{1.15}
\setlength{\tabcolsep}{5.0pt}
\begin{tabular}{p{1.8cm}p{3.8cm}cp{4.2cm}}
\toprule
\textbf{Category} & \textbf{Representative issue} & \textbf{Occurrence} & \textbf{Detection and status} \\
\midrule
Contraction syntax
& Inconsistent einsum labels or index routing
& 2
& Detected by static analysis, corrected automatically. \\

Array structure
& Shape handling in nonlocal displaced-field workflows
& 1
& Detected by static analysis, corrected automatically. \\

Interface consistency
& Output-format or variable-naming deviation
& 2
& Detected by critique, corrected automatically or non-impacting. \\

Runtime handling
& Gauge-context lifetime or execution-state error
& 1
& Detected during test execution, corrected automatically. \\

Convention mismatch
& Sequential-propagator sign or phase convention
& 2
& Detected by numerical comparison and analysis of the results. \\
\bottomrule
\end{tabular}
\end{table}


\section{Discussion}
\label{sec:discussion}

\textsc{LQCDMaster} demonstrates that a domain-specialized scientific computing agent can generate and execute reliable lattice QCD computing workflows according to natural-language research tasks. The present work also clarifies the reliability requirements for agentic scientific computing. The value of \textsc{LQCDMaster} lies in combining domain-specialized generation with explicit validation.

\textbf{Benchmark.} The scientific computing tasks in LQCD go beyond pure code generation, further requiring the capabilities of Wick contractions, spin--color algebra, gauge-link transport, source--sink kinematics, and phase conventions choices. To provide comprehensive and rigorous evaluation on LQCD scientific computing capabilities, we construct a 70-task benchmark spanning five
observable classes, to test whether an agent can reproduce the end-to-end scientific computing workflows of computing in frontier lattice QCD research. 

\textbf{Accuracy.} Across five representative observable classes, \textsc{LQCDMaster} reproduced expert implementations on 63 of 70 tasks at machine precision. Part of the remaining cases were traced to fixed convention mismatches associated with $\gamma_5$-hermiticity, rather than algebraic or numerical errors. The impressive accuracy of \textsc{LQCDMaster} is enabled by an architecture that integrates domain knowledge with robust executive workflow. The expert-annotated skills and deterministic tools constrain the brittle components (contraction structures, gauge-link paths, conventions, etc.) in LQCD computing, avoiding physically plausible but incorrect results. 

\textbf{Efficiency.} For real LQCD research, the main impact of \textsc{LQCDMaster} is the reduction of the marginal implementation burden, which accelerates the translation of an observable definition into reliable computational code. Tasks that typically require several hours of expert coding, debugging and convention checking can be generated and tested within minutes. This is particularly valuable for observables with dense implementation structures, such as nonlocal correlators, sequential-source three-point functions, baryonic contractions and Wilson-loop geometries, where manual errors are common and difficult to diagnose.

\textbf{New Paradigm.} New-observable exploration further demonstrates that domain-specialized agents can accelerate and automate scientific computing and assist real scientific research. The physics insights and research motivations came from physicists. Meanwhile, \textsc{LQCDMaster} helps reduce the engineering burden of implementing non-standard but well-defined observables by executing workflows for diagonal Wilson-line LCDA calculations and multi-hadron spectroscopy. Thus, we pioneer a new paradigm of human--AI scientific collaboration by lowering the implementation barrier for testing theoretically motivated ideas, such as alternative Wilson-line geometries, extended operator bases, displacement patterns, and contraction topologies.

\textbf{Limitations.} The current benchmark covers representative but not exhaustive LQCD computing tasks. More challenging cases, including disconnected diagrams, four-point or more-point correlations, finite-temperature observables and large multi-particle operators remain to be systematically tested. More advanced tasks and fine-grained validation metrics can be developed. Meanwhile, further work should expand the agentic capabilities such as automated fitting or physical interpretation pipelines, deterministic contraction generation, and more comprehensive skills. Combining more rigorous evaluation and broader agentic capabilities, our system can evolve from code-generation assistance toward an autonomous infrastructure for AI-assisted LQCD research.

\section{Methods: Agentic Workflow, Specialized Tools and Skills}
\label{sec:architecture}

\subsection{Agentic Workflow}
\textsc{LQCDMaster} is a tool-augmented, skill-guided, domain-specialized scientific computing agent that turns a natural-language LQCD research request into reviewable PyQUDA code. The system grounds generation in research workflows distilled from real LQCD tasks and is augmented with expert-annotated domain skills and a deterministic Einstein sum generation tool.

\begin{figure}[htbp]
  \centering
\includegraphics[width=1\textwidth]{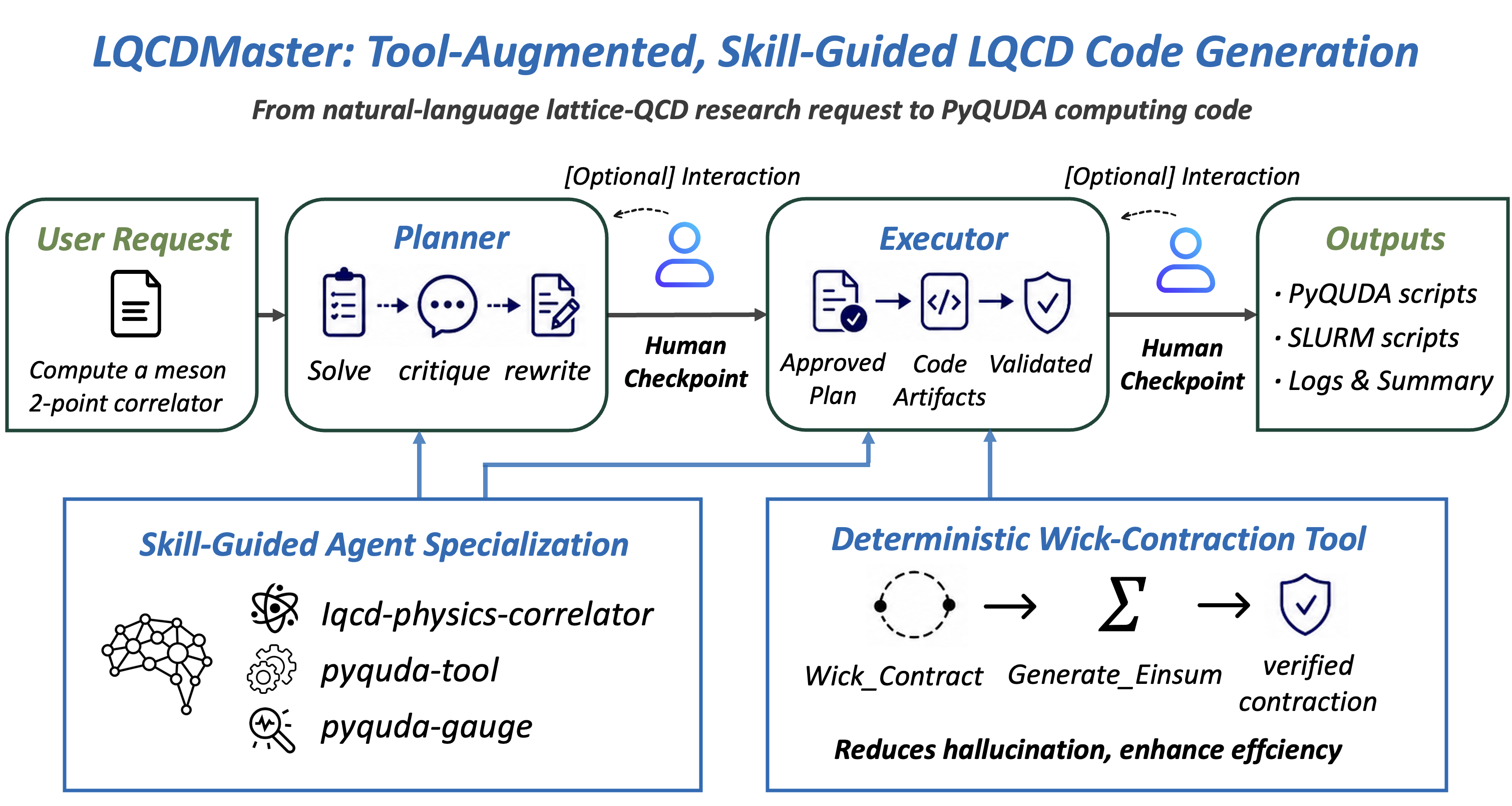}
\caption{Agentic workflow of \textsc{LQCDMaster}. The planner converts natural-language LQCD tasks into a structured scientific plan; the executor turns the approved plan into PyQUDA and SLURM artifacts; expert-annotated domain skills and a deterministic Wick-contraction tool enhance reliability and efficiency.}
  \label{fig:architecture}
\end{figure}

As shown in Fig.~\ref{fig:architecture}, \textsc{LQCDMaster} decomposes LQCD code generation into two stages, separating planning from execution.

\textbf{Planning stage}: The plan specifies the ensemble, observables, hadrons, propagators, correlator measurements, and expected output format. The planner agent operates in a solve--critique--rewrite loop: it conducts physical analysis and proposes a detailed technical plan; an independent critique checks whether the observables, propagators, and settings are reasonable and consistent. If any flaw exists, the plan will be  revised according to the critique.

Before the plan is handed to the executor, a human checkpoint allows physicists to review the proposed plan and engage in optional interactions to refine it. This acknowledges the current practical limitation that agent systems cannot yet guarantee absolute correctness for all LQCD observables.

\textbf{Execution stage}: The executor generates code according to the approved plan, producing PyQUDA and SLURM submission scripts. This stage is guided by PyQUDA-related skills that ensure code reliability by preventing invented API usage. Automated verification includes static analysis and LLM critique to catch errors before submission.

Agents have distinct core responsibilities in the two stages and produce corresponding artifacts, as illustrated in Table~\ref{tab:agent_workflow}.

\begin{table}[htbp]
  \centering
  \caption{\textsc{LQCDMaster} workflow artifacts.}
  \label{tab:agent_workflow}
  \begin{tabularx}{\textwidth}{lXX}
  \toprule
  \textbf{Stage} & \textbf{Agent role} & \textbf{Artifacts} \\
  \midrule
  Planning & Expose the observable, operators, propagators, kinematics, and ensemble assumptions &
  Computation plan and concise summary \\
  Execution & Realize the approved computation plan, generate the PyQUDA and SLURM scripts, and conduct tests to verify and revise the scripts  &
  Reliable PyQUDA script, test script, and full-run script \\
  \bottomrule
  \end{tabularx}
\end{table}

\subsection{The \texttt{generate\_einsum} Tool}
\label{sec:generate_einsum}

The primary specialized tool in \textsc{LQCDMaster} is \texttt{generate\_einsum}, which exposes a deterministic engine to the executor, as illustrated in Table~\ref{tab:generate_einsum} . It is designed to remove the most algebraically brittle part of correlator code generation---the contraction strings---from the language model's free-form output space. LQCD contraction strings encode flavor pairing, fermion signs, gamma insertions, epsilon tensors, and spin-color index ordering. A single wrong index or sign can yield code that executes successfully while measuring the wrong observable.

Instead of writing these scripts directly, the executor requests a contraction in physics terms (e.g., meson two-point, baryon two-point, baryon three-point sequential-source sink block). \texttt{generate\_einsum} then enumerates the valid Wick pairings, applies sign and topology conventions, constructs the spin-color labels, and returns a ready-to-insert contraction specification. The generated PyQUDA script contains the resulting einsum strings; it does not depend on \texttt{generate\_einsum} at runtime.

This separation changes the failure mode of the system. The model remains responsible for high-level scientific choices that require judgment---the observable, source-sink setup, current insertion, and solver strategy. Exact symbolic construction, however, is delegated to deterministic code. \texttt{generate\_einsum} is therefore a reliability boundary between physical planning and algebraic realization.

\begin{table}[htbp]
  \centering
  \caption{Contraction families exposed through \texttt{generate\_einsum}.}
  \label{tab:generate_einsum}
  \begin{tabularx}{\textwidth}{lX}
  \toprule
  Type & Role in generated LQCD code \\
  \midrule 
  \texttt{meson\_2pt} & Meson two-point contractions with explicit source and sink gamma structures \\
  \texttt{baryon\_2pt} & Baryon two-point contractions with epsilon tensors, projector, and topology signs \\
  \texttt{multi\_hadron\_2pt} & Multi-hadron two-point contractions \\
  \texttt{meson\_3pt} & Sequential-source sink block for meson three-point functions \\
  \texttt{baryon\_3pt} & Sequential-source sink block for baryon three-point functions\\
  \bottomrule
  \end{tabularx}
\end{table}

\subsection{Skill-Guided Agent Specialization}
\label{sec:skills}

To enhance reliability and efficiency, \textsc{LQCDMaster} is augmented with expert-annotated domain skills. Each skill is a compact prior over a class of LQCD tasks: how to translate an observable into operators and propagators, how to implement the computation in PyQUDA, or how to handle gauge-only measurements.

Skills are stage-routed: the planner is equipped with correlator physics skills (observables, interpolating operators, Wick-contraction structure, momentum projection, sequential-source requirements); the executor receives PyQUDA skills (source construction, solver parameters, MPI data layout, contraction implementation, output formatting); and gauge-only tasks are handled by a dedicated \texttt{pyquda-gauge} skill.

The coupling of domain skills and tools makes domain routing operational, aligning the agent's knowledge with the tools valid for the selected physics workflow.

\begin{table}[htbp]
  \centering
  \caption{Main skills in \textsc{LQCDMaster}.}
  \label{tab:skills}
  \begin{tabularx}{\textwidth}{lXc}
  \toprule
  Skill & Content & Agent stage \\
  \midrule
  \texttt{lqcd-physics-correlator} & Hadronic operators, Wick-contraction
  conventions, flavor assignments, and $\gamma_5$-hermiticity & Planner \\
  \texttt{pyquda-tool} & PyQUDA implementation patterns, solver setup,
  contraction use, MPI gather, and common code-generation mistakes & Executor \\
  \texttt{pyquda-gauge} & Gauge loading, Wilson loops, Polyakov loops, and
  gauge-only measurement conventions & Planner/Executor \\
  \bottomrule
  \end{tabularx}
\end{table}

\section{Data and Code Availability}

All source code for \textsc{LQCDMaster} is available at \url{https://github.com/sjtu-sai-agents/LQCD_Master}, and includes: 

\begin{itemize}
    \item The core architecture, domain-specialized skills and the \texttt{generate\_einsum} tool of \textsc{LQCDMaster}.
    \item The LQCD benchmark with 70 scientific computing tasks.
    \item Generated PyQUDA measurement scripts, SLURM submission scripts, and production output data in experiments.
\end{itemize}

\section*{Acknowledgements}
We thank Feng Xu, Zhaofeng Liu, Yu Meng, and Yibo Yang for valuable discussions regarding manual coding effort involved in constructing the benchmark. This work is supported in part by National Natural Science Foundation of China under grants No. 12125503, 12305103.  

\begin{appendices}
\section{Architecture Robustness Validation with DeepSeek-V4-Pro}

A central question for LLM-based scientific agents is whether their reliability is tied to a particular foundation model. To assess this dependence, we repeated the full 70-task benchmark using DeepSeek-V4-Pro as the backbone LLM, while keeping the \textsc{LQCDMaster} architecture, expert skill files, tool interfaces and solve--critique--rewrite pipeline unchanged. Thus, only the chat-completion endpoint was replaced.

Table~\ref{tab:cross_validation_ds} summarizes the cross-backbone validation results. With DeepSeek-V4-Pro as backbone, the validation achieved 56/70 exact matches at machine precision, corresponding to an accuracy of 80.0\%. Three additional cases exhibited clean global sign mismatches consistent with fixed convention choices. Wilson loops reached 100\% exact agreement, while local 2pt and baryon 3pt functions achieved 80.0\% and 86.7\%, respectively. 

The remaining discrepancies were concentrated in nonlocal 2pt and meson 3pt functions. Manual inspection showed that most nonlocal 2pt mismatches arose from output-layout inconsistencies rather than incorrect correlator values, whereas meson 3pt failures were mainly caused by incorrect quark-flavor assignments in model-generated task parameters. These errors identify the limit of the current system: syntactic and algebraic constraints are well controlled by the tool layer, but semantic validation of observable-specific task parameters remains partially dependent on the backbone model.

Notably, among the challenging 3pt tasks, DeepSeek-V4-Pro produced more failures for meson than for baryon observables, differing from the error distribution observed with GPT-5.4. We speculate that this difference may reflect weaker domain-specific knowledge of observable-dependent flavor assignments acquired during model training.

This result further suggests that, the architecture of \textsc{LQCDMaster} provides substantial robustness across backbone models, while the residual failures remain sensitive to the in-domain knowledge of the backbone LLMs.

\begin{table}[htbp]
\centering
\caption{
Architecture robustness validation of \textsc{LQCDMaster} with DeepSeek-V4-Pro as the backbone model.
}
\label{tab:cross_validation_ds}
\small
\renewcommand{\arraystretch}{1.15}
\setlength{\tabcolsep}{5.0pt}
\begin{tabular}{lccccc}
\toprule
\textbf{Observables} & \textbf{Tasks} & \textbf{Matched} & \textbf{Convention Mismatch} & \textbf{Failure} & \textbf{Accuracy} \\
\midrule
Local 2pt              & 20 & 16 & 1 & 3 & 80.0\%  \\
Nonlocal 2pt           & 10 & 8  & 0 & 2 & 80.0\%  \\
Wilson loop            & 12 & 12 & 0 & 0 & 100.0\% \\
Meson 3pt              & 13 & 7  & 1 & 5 & 53.8\%  \\
Baryon 3pt             & 15 & 13 & 1 & 1 & 86.7\% \\
\midrule
\textbf{Total}         & \textbf{70} & \textbf{56} & \textbf{3} & \textbf{11} & \textbf{80.0\%} \\
\bottomrule
\end{tabular}

\vspace{0.5em}
\begin{minipage}{0.95\linewidth}
\footnotesize
\begin{itemize}
\setlength{\itemsep}{0.2em}
\setlength{\leftmargin}{1.2em}
\item \textbf{Matched}: numerical agreement with the expert-written reference implementation at machine precision, $|\Delta|\lesssim10^{-12}$, under identical input data, normalization choices and output conventions.
\item \textbf{Convention mismatch}: minor discrepancy such as a single global sign or phase transformation, with the full correlator structure preserved after fixed convention alignment.
\item \textbf{Failure}: any remaining algebraic, numerical, formatting or execution-level discrepancy after convention alignment.
\end{itemize}
\end{minipage}
\end{table}

\section{Comparison with General Coding Agent}
To demonstrate the relative efficiency of our \textsc{LQCDMaster}, we further compared it with Claude Code, a representative general coding agent. In this baseline, Claude Code was provided the same benchmark tasks and access to PyQUDA documentation\cite{Jiang:2024lto} in the working space, but was not equipped with LQCD-specialized skills and tools.

As a result, \textbf{Claude Code failed to yield executable scripts for any of the 20 local two-point-function tasks}, the simplest class of observables in the benchmark.

These results indicate that guidebook alone is insufficient for reliable LQCD computing workflow generation. It is essential to integrate domain knowledge like domain-specialized skills and deterministic algebraic tools into the agentic scientific computing system for robust LQCD code generation.

\end{appendices}

\bibliographystyle{unsrt}

\end{document}